%% file: ms.tex
\documentclass[aps,prd,preprint,groupedaddress]{revtex4}

\usepackage{graphicx}
\usepackage{side}

\def\dfrac#1#2{\displaystyle\frac{#1}{#2}}
\newcommand{\ovl}[1]{\overline{#1}}

\newcommand{\p}{\partial}
\newcommand{\kslash}{k\kern-1ex /}
\newcommand{\pslash}{p\kern-1ex /}
\newcommand{\lslash}{l\kern-1ex /}
\newcommand{\sslash}{s\kern-1ex /}
\newcommand{\Dslash}{{\cal D}\kern-1.5ex /}
\newcommand{\bpsi}{\overline{\psi}}

\newcommand{\beqa}{\begin{eqnarray}}
\newcommand{\eeqa}{\end{eqnarray}}

\newcommand{\be}{\begin{equation}}
\newcommand{\ee}{\end{equation}}
\newcommand{\ben}{\begin{eqnarray}}
\newcommand{\een}{\end{eqnarray}}
\newcommand{\nn}{\nonumber}
\def\lsim{\raise0.3ex\hbox{$<$\kern-0.75em\raise-1.1ex\hbox{$\sim$}}}
\def\gsim{\raise0.3ex\hbox{$>$\kern-0.75em\raise-1.1ex\hbox{$\sim$}}}
\def\simgt{\rlap{\lower 3.5 pt\hbox{$\mathchar \sim$}}\raise 1pt \hbox {$>$}}
\def\simlt{\rlap{\lower 3.5 pt\hbox{$\mathchar \sim$}}\raise 1pt \hbox {$<$}}

\newcommand{\msbar}{{\overline {\rm MS}}}

\begin{document}


\title{The Lattice $\Lambda$ Parameter in Domain Wall QCD}


\author{Sinya Aoki$^a$ and Yoshinobu Kuramashi$^b$}
\affiliation{$^a$Institute of Physics, University of Tsukuba, 
Tsukuba, Ibaraki 305-8571, Japan \\
$^b$Institute of Particle and Nuclear Studies,
High Energy Accelerator Research Organization(KEK),
Tsukuba, Ibaraki 305-0801, Japan}


\date{\today}

\begin{abstract}
We evaluate the ratio of the scale parameter $\Lambda$ in
domain wall QCD to the one in the continuum theory  at one loop level
incorporating the effect of massless quarks.
We show that the Pauli-Villars regulator is required to subtract
the unphysical massive fermion modes which emerge in the fermion loop
contributions to the gluon self energy.
Detailed results are presented 
as a function of the domain wall height $M$.   
\end{abstract}


\maketitle


\section{Introduction}
\label{sec:intro}

The domain-wall fermion formalism provides us an opportunity
to realize the chiral symmetry at finite lattice spacing
with any number of quark flavors\cite{shamir,FS}.
Since it is well known that the chiral symmetry plays an essential role 
in low energy QCD, this point is a superior feature over the Wilson and 
the Kogut-Susskind quark actions to perform lattice QCD simulations.
While in the past two decade both conventional quark actions 
have been widely used to make unquenched QCD simulations incorporating 
the dynamical quark effects, recent progress in computational
technologies allows us to embark on a full QCD simulation with the   
domain-wall fermion\cite{dw_full}.

One primary step in the full QCD simulation is to determine 
the renormalized coupling constant, which is a fundamental parameter in QCD.
At the one-loop level the renormalized coupling 
in the continuum $\msbar$ scheme $g_\msbar(\mu)$ at the scale $\mu$ 
and that on the lattice $g_0$
are related by
\be
\dfrac{1}{g_\msbar^2(\mu )}
= \dfrac{1}{g_0^2} + \left(d_g +\dfrac{22}{16\pi^2} \log (\mu a)\right)
+N_f\left(d_f -\dfrac{4}{48\pi^2} \log (\mu a)\right)
\label{eq:g2}
\ee
where $d_g$ and $d_f$ are renormalization 
constants: the former depends on the gauge 
action and the latter on the quark action.
These constant terms determine 
the ratio of the lattice $\Lambda$ parameter
and the continuum one at the one-loop level. 
 From the expressions of the $\Lambda$ parameters,
\ben
\Lambda_L&=&\frac{1}{a}(\beta_0 g_0^2)^{-\beta_1/2\beta_0^2}
{\rm exp}\left(-\frac{1}{2\beta_0 g_0^2}\right),\\
\Lambda_\msbar&=&\mu(\beta_0 g_\msbar^2(\mu))^{-\beta_1/2\beta_0^2}
{\rm exp}\left(-\frac{1}{2\beta_0 g_\msbar^2(\mu)}\right)
\een
with
\ben
\beta_0&=&\frac{1}{16\pi^2}\left(11-\frac{2}{3}N_f\right),\\
\beta_1&=&\frac{1}{(16\pi^2)^2}\left(102-\frac{38}{3}N_f\right),
\een
we obtain the ratio
\ben
\frac{\Lambda_\msbar}{\Lambda_L}&=&(a\mu ){\rm exp}
\left(-\frac{1}{2\beta_0}\left\{\frac{1}{g_\msbar^2(\mu)}-
\frac{1}{g_0^2}\right\}\right)+O(g_0^2),\\
&=&{\rm exp}\left(-\frac{1}{2\beta_0}
\left\{d_g+N_f d_f\right\}\right)+O(g_0^2).
\label{eq:ratio}
\een
Thus the finite renormalization between the coupling constants
defined by different regularization schemes is 
to relate their $\Lambda$ parameters.
In Table~\ref{tab:MF} we list the renormalization constant 
$d_g$, which has been evaluated 
for various gauge actions\cite{dg_plaq,Weisz83,dg_w,dg_iwa,dg_dbw2}.

In this paper we make a perturbative determination of $d_f$ for the
domain-wall fermion, which is a necessary ingredient to extract the
renormalized coupling constant 
from the numerical simulation with domain-wall QCD.
While the perturbative techniques are well developed in our previous
publications\cite{pt_at,pt_awi,pt_2,pt_34,pt_pngn,pt_rg}, a characteristic
feature in this calculation is to require the Pauli-Villars field 
to subtract the unphysical massive fermion modes, 
which emerge from the fermion loop diagrams.
  
This paper is organized as follows. In Sec.~II we introduce the 
domain-wall fermion action with the Pauli-Villars regulator.
We give their Feynman rules relevant for the
present calculation. In Sec.~III we evaluate the quark loop contributions
to the gluon self energy in the continuum and on the lattice.
The ratio of continuum and lattice $\Lambda$ parameters is given 
in Sec.~IV. Our conclusions are summarized in Sec.~V. 

The physical quantities are expressed in lattice units and 
the lattice spacing $a$ is suppressed unless necessary.
We take  SU($N_c$) gauge group with the gauge coupling constant $g$.

\section{Action and Feynman Rules}

We explain the domain-wall fermion action accompanied with its Pauli-Villars
regulator in the present calculation.
For the fermion action we take Shamir's form\cite{shamir}:
\ben
S_{\rm DW} &=&
\sum_{n} \sum_{s=1}^{N_s} \Biggl[ \frac{1}{2} \sum_\mu
\left( \bpsi(n)_s (-r+\gamma_\mu) U_\mu(n) \psi(n+{\hat \mu})_s
+ \bpsi(n)_s (-r-\gamma_\mu) U_\mu^\dagger(n-{\hat \mu}) 
\psi(n-{\hat \mu})_s \right)
\nn\\&&
+ \frac{1}{2}
\left( \bpsi(n)_s (1+\gamma_5) \psi(n)_{s+1}
+ \bpsi(n)_s (1-\gamma_5) \psi(n)_{s-1} \right)
+ (M-1+4r) \bpsi(n)_s \psi(n)_s \Biggr]
\nn\\&&+
 m \sum_n \left( \bpsi(n)_{N_s} P_{R} \psi(n)_{1}
+ \bpsi(n)_{1} P_{L} \psi(n)_{N_s} \right),
\label{eq:action_q}
\een
where $U_\mu$ is the link variable of the SU($N_c$) gauge group.
We label a four-dimensional space-time coordinate by $n$ and an extra 
fifth-dimensional one by $s$, which runs from $1$ to $N_s$.
Since no gauge interaction is imposed along 
the fifth dimension, we are also allowed to consider that the index $s$ labels
the $N_s$ ``flavor'' space.
The parameter $m$ denotes the physical quark mass and at $m=0$ 
one chiral zero mode is realized under the condition
$0 < M < 2$ for the Dirac ``mass'' (domain-wall height) $M$. 
The Wilson parameter is set to $r=-1$.
$P_{R/L}$ are projection operators defined by $P_{R/L}=(1\pm\gamma_5)/2$.

The ``physical'' quark fields on the four-dimensional space-time 
are constructed in terms of the fermion fields at the boundaries,
\ben
q(n) = P_R \psi(n)_1 + P_L \psi(n)_{N_s},\\
\ovl{q}(n) = \bpsi(n)_{N_s} P_R + \bpsi(n)_1 P_L,
\label{eq:quark}
\een
with which we can extract the zero mode of domain-wall fermion at $m=0$.

We also introduce the Pauli-Villars field to subtract
the massive modes of the domain-wall fermion, which emerge from the 
quark loop diagram in proportion to $N_s$ in the large $N_s$ limit.
The action is given by
\be
S_{\rm PV}=S_{\rm DW}(\bpsi\rightarrow \bar\phi,\psi\rightarrow \phi,m=-1),
\ee
where $\phi$ is the $N_s$-flavor Wilson-Dirac boson, satisfying
the anti-periodic boundary condition in the fifth direction
because of the special value $m=-1$\cite{vranas}. 
Here we have to remark one point:
no Pauli-Villars field has been introduced
in the previous calculations for the renormalization factors 
of the composite operators\cite{pt_at,pt_awi,pt_2,pt_34,pt_pngn,pt_rg},
since it requires no quark loop diagram. 


For the gauge action we employ the following form:
\be
S_{\rm gluon} = \frac{1}{g^2}\left\{
c_0 \sum_{\rm plaquette} {\rm Tr}U_{pl}
+ c_1  \sum_{\rm rectangle} {\rm Tr} U_{rtg}
+ c_2 \sum_{\rm chair} {\rm Tr} U_{chr}
+ c_3 \sum_{\rm parallelogram} {\rm Tr} U_{plg}\right\}, 
\label{eqn:RG}
\ee
where the first term represents the standard plaquette action, and the 
remaining terms are six-link loops formed by a $1\times 2$ rectangle, 
a bent $1\times 2$ rectangle (chair) and a 3-dimensional parallelogram. 
The coefficients $c_0, \cdots, c_3$ satisfy the normalization condition
\be
c_0+8c_1+16c_2+8c_3=1. 
\ee
There are several choices for these parameters:
$c_1=c_2=c_3=0$(Plaquette),
$c_1=-1/12$, $c_2=c_3=0$(Symanzik)\cite{Weisz83,LW} 
$c_1=-0.331$, $c_2=c_3=0$(Iwasaki), $c_1=-0.27, 
c_2+c_3=-0.04$(Iwasaki')
\cite{Iwasaki83}, $c_1=-0.252, c_2+c_3=-0.17$(Wilson)\cite{Wilson}
and $c_1 = -1.40686$, $c_2=c_3=0$(DBW2)\cite{qcdtaro}.
The last four cases are called the RG-improved gauge action, whose 
parameters are chosen to be the values suggested by 
some approximate renormalization group analyses.
Some of these actions are now getting widely used to simulate 
the domain-wall QCD\cite{cppacs_rg,rbc_rg}, since
they are expected to realize smooth gauge field fluctuations 
which approximate the continuum gauge field 
better than the naive plaquette action.

To prepare the Feynman rule we develop weak-coupling perturbation theory 
by writing the link variable in terms of the gauge potential
\be
U_\mu(n)=\exp \left(igaT^A A_\mu^A\left(n+\frac{1}{2}\hat\mu\right)\right),
\ee
where $T^A$ ($A=1,\dots,N^2_c-1$) is a generator of color SU($N_c$).
Here we omit the expression of the gluon propagator and vertices, 
since they are not relevant for the present calculation.
Quark-gluon vertices are identical to those in 
the $N_s$ flavor Wilson fermion. 
We need one- and two-gluon vertices for the present calculation:
\ben
V_{1\mu}^A (q,p)_{st}
&=& -g T^A \{i \gamma_\mu \cos (q_\mu/2 + p_\mu/2)
  + r \sin (q_\mu/2 + p_\mu/2) \} \delta_{st},\\
&=& -g T^A v_{1\mu} (q,p)\delta_{st},
\\
V_{2\mu\nu}^{AB} (q,p)_{st}
&=& \frac{a}{2} g^2 \frac{1}{2} \{T^{A}, T^{B}\}
\{ i \gamma_\mu \sin (q_\mu/2 + p_\mu/2)
-r \cos (q_\mu/2 + p_\mu/2) \}\delta_{\mu\nu} \delta_{st},\\
&=& \frac{a}{2} g^2 \frac{1}{2} \{T^{A}, T^{B}\}
v_{2\mu\nu} (q,p)\delta_{st},
\een
where $p$ and $q$ represent incoming and outgoing quark 
momentum, respectively. 

The fermion propagator is obtained by inverting the domain-wall Dirac operator.
In the present one-loop calculation we will 
take $N_s\rightarrow\infty$ to neglect the contribution of 
the $O(e^{-\alpha N_s})$ terms with the positive constant $\alpha$ given
in eq.(\ref{eq:alpha}).
To this end it is sufficient to consider 
the following component of the fermion propagator:
\ben
S_F(p)_{st} &\equiv& \langle \psi(-p)\bpsi(p)\rangle_{st}\\
&=&
\left( -i\gamma_\mu \sin p_\mu + W^- \right)_{su} G_R(u,t) P_+
\nn\\&&+
\left( -i\gamma_\mu \sin p_\mu + W^+ \right)_{su} G_L(u,t) P_-,
\een
where the sum over the same index is taken implicitly.
$W^{+/-}$ and $G_{R/L}$ are given by 
\ben
&&
W^{+} (p)_{st} =
\pmatrix{
-W(p) & 1     &        &       \cr
      & -W(p) & \ddots &       \cr
      &       & \ddots & 1     \cr
m     &       &        & -W(p) \cr
},
\label{eqn:massm_p}
\\&&
W^{-} (p)_{st} =
\pmatrix{
-W(p) &        &        & m     \cr
1     & -W(p)  &        &       \cr
      & \ddots & \ddots &       \cr
      &        & 1      & -W(p) \cr
},
\label{eqn:massm_m}
\\&&
W(p) = 1-M -r \sum_\mu (1-\cos p_\mu).
\\
G_{R} (s, t) 
&=&
\frac{A}{F}
\Bigl[
-(1-m^2) \left(1-W e^{-\alpha}\right) e^{\alpha (-2N_s+s+t)}
-(1-m^2) \left(1-W e^\alpha\right) e^{-\alpha (s+t)}
\nn\\&&
-2W \sinh (\alpha) m
 \left( e^{\alpha (-N_s+s-t)} + e^{\alpha (-N_s-s+t)} \right)
\Bigr]
+ A e^{-\alpha |s-t|},
\label{eq:GR}
\\
G_{L} (s, t) 
&=&
\frac{A}{F}
\Bigl[
-(1-m^2) \left(1-We^{\alpha}\right) e^{\alpha (-2N_s+s+t-2)}
-(1-m^2) \left(1-We^{-\alpha}\right) e^{\alpha (-s-t+2)}
\nn\\&&
-2W \sinh (\alpha) m
 \left( e^{\alpha (-N_s+s-t)} +  e^{\alpha (-N_s-s+t)} \right)
\Bigr]
+ A e^{-\alpha |s-t|},
\label{eq:GL}
\\
\cosh (\alpha) &=& \frac{1+W^2+\sum_\mu \sin^2 p_\mu}{2|W|},
\label{eq:alpha}
\\
A&=& \frac{1}{2W \sinh (\alpha)},
\label{eq:A}
\\
F &=& 1-e^{\alpha} W-m^2 \left(1-W e^{-\alpha}\right).
\label{eq:F}
\een
In case that $W$ becomes negative the fermion propagator is given with
the replacement $e^{\pm\alpha}\to-e^{\pm\alpha}$.
In the present calculation we consider only $m=0$ case.
The Feynman rule for the Pauli-Villars regulator is obtained by
setting the quark mass to $m=-1$.

\section{Fermion loop contribution to the gluon self-energy}

\begin{figure}[h]
\centering{
\hskip -0.0cm
\includegraphics[width=140mm,angle=0]{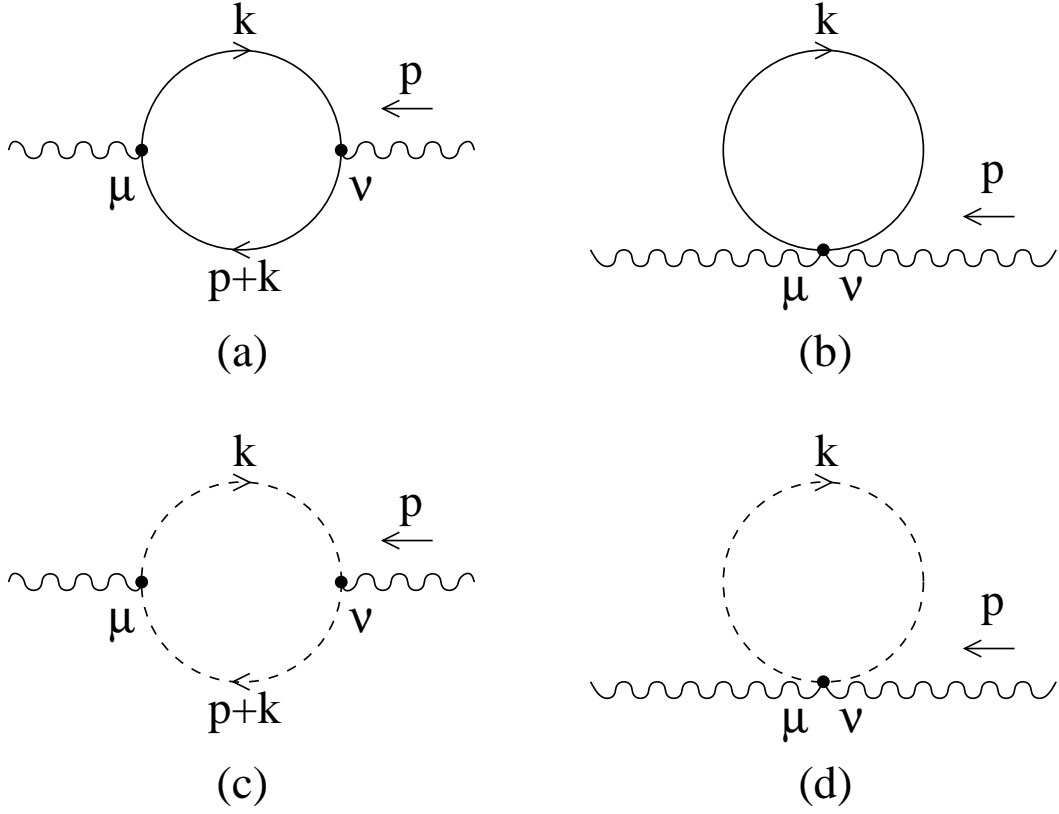}     
}
\caption{One loop corrections to the gluon self energy.
Solid line is for the propagator of the domain-wall fermion 
and broken one for the Pauli-Villars regulator.}
\label{fig:vp}
\vspace{8mm}
\end{figure}

We first present the calculation of the vacuum polarization diagram
in Fig.~\ref{fig:vp} (a) employing a continuum regularization scheme
with the massless quarks.
Here we choose the naive dimensional regularization (NDR) 
as the ultraviolet regularization
scheme. Both the Dirac matrices and the fermion loop momenta 
of the Feynman integrals 
are defined in $D=4-2\epsilon$ ($\epsilon>0$) in this scheme.
To regularize the infrared divergence we take the external 
gluon momentum $p$ to be off-shell $p^2\ne 0$.
   
The gluon two point function in Fig.~\ref{fig:vp} (a) is expressed as
\be
\Pi_{\mu\nu}^{AB}(p)=-N_f\int\frac{d^Dk}{(2\pi)^D}{\rm Tr}
\left[(-g)i\gamma_\mu T_{ab}^{A}\frac{1}{i(\pslash+\kslash)}
(-g)i\gamma_\nu T_{ba}^{B}\frac{1}{i\kslash}\right].
\ee
After some algebra we obtain
\be
\Pi_{\mu\nu}^{AB}(p)=-N_f\frac{g_0^2}{(4\pi)^2}\delta_{AB}
(\delta_{\mu\nu}p^2-p_\mu p_\nu)
\left\{\frac{2}{3}\frac{1}{\bar\epsilon}
+\frac{2}{3}\log(\mu^2/p^2)+\Delta_{\rm cont}\right\}
\label{eq:vp_cont}
\ee
with
\ben
\Delta_{\rm cont} = \frac{10}{9} 
\een
where $1/\bar\epsilon=1/\epsilon-\gamma+\ln(4\pi)$ and $\mu$ is
the renormalization scale.
In the $\msbar$ scheme, the pole term $1/\bar\epsilon$ 
should be eliminated.


Let us turn to the calculation of the one loop corrections to the
gluon self energy on the lattice.
We illustrate the relevant diagrams in Fig.~\ref{fig:vp}:
(a), (b) for the contribution of the domain wall fermion  
and (c), (d) for the Pauli-Villars regulators.
Each loop diagram is expressed by
\ben
{\Pi_{\mu\nu}^{(a)}}^{AB}(p)&=&-N_f\int^{\pi}_{-\pi}
\frac{d^4k}{(2\pi)^4}{\rm Tr}
\left[V_{1\mu}^A(k,p+k)_{uv}S_F(p+k)_{vs}V_{1\nu}^B(p+k,k)_{st}
S_F(k)_{tu}\right],\\
{\Pi_{\mu\nu}^{(b)}}^{AB}(p)&=&-N_f\int^{\pi}_{-\pi}
\frac{d^4k}{(2\pi)^4}{\rm Tr}
\left[V_{2\mu\nu}^{AB}(k,k)_{st}S_F(k)_{ts}\right],\\
{\Pi_{\mu\nu}^{(c)}}^{AB}(p)&=&+N_f\int^{\pi}_{-\pi}
\frac{d^4k}{(2\pi)^4}{\rm Tr}
\left[{\ovl V}_{1\mu}^A(k,p+k)_{uv}{\ovl S}_F(p+k)_{vs}
{\ovl V}_{1\nu}^B(p+k,k)_{st}
{\ovl S}_F(k)_{tu}\right],\\
{\Pi_{\mu\nu}^{(d)}}^{AB}(p)&=&+N_f\int^{\pi}_{-\pi}
\frac{d^4k}{(2\pi)^4}{\rm Tr}
\left[{\ovl V}_{2\mu\nu}^{AB}(k,k)_{st}{\ovl S}_F(k)_{ts}\right],
\een
where ${\ovl V}_{1\mu}^A$, ${\ovl V}_{2\mu\nu}^A$ and ${\ovl S}_F$
denote the one-, two-gluon vertices and the propagator 
for the Pauli-Villars regulators.
The sum of these contributions is expressed by $\Pi_{\mu\nu}^{AB}(p)$: 
\ben
\Pi_{\mu\nu}^{AB}(p)&=&{\Pi_{\mu\nu}^{(a)}}^{AB}(p)
+{\Pi_{\mu\nu}^{(b)}}^{AB}(p)
+{\Pi_{\mu\nu}^{(c)}}^{AB}(p)+{\Pi_{\mu\nu}^{(d)}}^{AB}(p)\\
&=& \int_{-\pi}^{\pi}\frac{d^4 k}{(2\pi)^4} I_{\mu\nu}^{AB}(k,p).
\een
Since the function $\Pi_{\mu\nu}^{AB}(p)$ has an infrared divergence
for $p^2\rightarrow 0$, we extract it by employing an analytically integrable
expression ${\tilde I}_{\mu\nu}^{AB}(p)$ which has the same infrared
behavior as $I_{\mu\nu}^{AB}(p)$,
\ben  
\Pi_{\mu\nu}^{AB}(p) &=& \int_{-\pi}^{\pi}\frac{d^4 k}{(2\pi)^4}
\tilde I_{\mu\nu}^{AB}(k,p) 
+ \int_{-\pi}^{\pi}\frac{d^4 k}{(2\pi)^4} 
\left\{I_{\mu\nu}^{AB}(k,p)-\tilde I_{\mu\nu}^{AB}(k,p)\right\},
\label{eq:Pisubtaction}
\een
where
\be
\tilde I_{\mu\nu}^{AB}(k,p)=
\theta(\Lambda^2-k^2){\rm Tr}
\left[(-g)i\gamma_\mu T_{ab}^{A}\frac{1}{i(\pslash+\kslash)}
(-g)i\gamma_\nu T_{ba}^{B}\frac{1}{i\kslash}\right],
\ee
with  a cut-off $\Lambda$ $(\le\pi)$.
Now we find that the second term in the right-hand side 
of eq.(\ref{eq:Pisubtaction}) is
regular in terms of $p$, we can make a Taylor expansion,
\ben
\Pi_{\mu\nu}^{AB}(p) 
=&& \int_{-\pi}^{\pi}\frac{d^4 k}{(2\pi)^4} \tilde I_{\mu\nu}^{AB}(k,p)\nn\\ 
&+&\int_{-\pi}^{\pi}\frac{d^4 k}{(2\pi)^4} 
\left\{I_{\mu\nu}^{AB}(k,0)-\tilde I_{\mu\nu}^{AB}(k,0)\right\} \nn\\ 
&+& \int_{-\pi}^{\pi}\frac{d^4 k}{(2\pi)^4}
p_\rho \frac{\p}{\p p_\rho}
\left\{I_{\mu\nu}^{AB}(k,p)
-\tilde I_{\mu\nu}^{AB}(k,p)\right\}\vert_{p=0} \nn\\
&+& \int_{-\pi}^{\pi}\frac{d^4 k}{(2\pi)^4}
\frac{1}{2} p_\rho p_\lambda \frac{\p^2}{\p p_\rho \p p_\lambda}
\left\{I_{\mu\nu}^{AB}(k,p)-\tilde I_{\mu\nu}^{AB}(k,p)\right\}\vert_{p=0},
\label{eq:Piexpansion}
\een
where $\rho,\lambda=1,2,3,4$.

Let us consider each term in eq.(\ref{eq:Piexpansion}).
We first point out that the linear term in $p$ is not allowed
from the Lorentz symmetry.
The integration of 
$\tilde I_{\mu\nu}^{AB}(k,p)-\tilde I_{\mu\nu}^{AB}(k,0)$ is easily 
performed analytically, 
\ben
&&\int_{-\pi}^{\pi}\frac{d^4 k}{(2\pi)^4}
\left\{\tilde I_{\mu\nu}^{AB}(k,p)-\tilde I_{\mu\nu}^{AB}(k,0)\right\}\nn\\ 
&=& -N_f\frac{g_0^2}{(4\pi)^2}\delta_{AB}\left[
(\delta_{\mu\nu}p^2-p_\mu p_\nu)
\left\{\frac{2}{3}\log(\Lambda^2/p^2)+\frac{5}{9}\right\}+\delta_{\mu\nu}p^2/3 
\right].
\label{eq:int_IR}
\een
As for $\Pi_{\mu\nu}^{AB}(0) =\int_{-\pi}^{\pi} I_{\mu\nu}^{AB}(k,0)$,
we find that this function vanishes.
This can be shown as follows.
For the fermion contribution,
\ben
{\Pi_{\mu\nu}^{(a)}}^{AB}(0)&=&+N_f\frac{g_0^2}{2}\delta_{AB}\int^{\pi}_{-\pi}
\frac{d^4k}{(2\pi)^4}{\rm Tr}
\left[v_{1\mu}(k,k)S_F(k)_{ts}v_{1\nu}(k,k)
S_F(k)_{st}\right],\\
&=&-N_f\frac{g_0^2}{2}\delta_{AB}\int^{\pi}_{-\pi}
\frac{d^4k}{(2\pi)^4}{\rm Tr}
\left[v_{1\mu}(k,k)\frac{\partial}{\partial k_\nu}S_F(k)_{tt}\right],\\
&=&+N_f\frac{g_0^2}{2}\delta_{AB}\int^{\pi}_{-\pi}
\frac{d^4k}{(2\pi)^4}{\rm Tr}
\left[\frac{\partial}{\partial k_\nu}v_{1\mu}(k,k)S_F(k)_{tt}\right],\\
&=&-{\Pi_{\mu\nu}^{(b)}}^{AB}(0),
\een
where we use the identities
\ben
&&\frac{\partial}{\partial k_\nu}S_F^{-1}(k)_{st}=v_{1\nu}(k,k)\delta_{st},\\
&&\frac{\partial}{\partial k_\nu}[S_F^{-1}(k)_{st}S_F(k)_{tu}]=0,\\
&&\frac{\partial}{\partial k_\nu}v_{1\mu}(k,k)=v_{2\mu\nu}(k,k).
\een
In a similar way we can show 
\be
{\Pi_{\mu\nu}^{(c)}}^{AB}(0)=-{\Pi_{\mu\nu}^{(d)}}^{AB}(0)
\ee
for the contribution of the Pauli-Villars regulator.
After all, the sum of all the contributions vanishes: 
\be
\Pi_{\mu\nu}^{AB}(0)={\Pi_{\mu\nu}^{(a)}}^{AB}(0)+{\Pi_{\mu\nu}^{(b)}}^{AB}(0)+{\Pi_{\mu\nu}^{(c)}}^{AB}(0)+{\Pi_{\mu\nu}^{(d)}}^{AB}(0)=0. 
\ee

We now turn to the last term in eq.~(\ref{eq:Piexpansion}).
The calculation of this term requires tedious algebra and yields
quite lengthy results that make it practically difficult 
to give their full expression here.
However, we think that it would be instructive 
to explain why the Pauli-Villars regulator is required by  
showing a part of the calculation. 
Let us take the term proportional to
${\rm Tr}[\gamma_\mu\gamma_\alpha\gamma_\nu\gamma_\beta]$ 
as an example, 
\ben
{I_{\mu\nu}^{(4\gamma)}}^{AB}(k,p)&=&
-N_f g_0^2\frac{\delta_{AB}}{2}
{\rm Tr}[\gamma_\mu\gamma_\alpha\gamma_\nu\gamma_\beta]
{\rm cos}(k_\mu+p_\mu/2){\rm sin}(k_\alpha+p_\alpha)
{\rm cos}(k_\nu+p_\nu/2){\rm sin}(k_\beta)\nn\\
&&\times\left\{G_R^\prime(s,t)G_R(t,s)
-{\ovl G}_R^\prime(s,t){\ovl G}_R(t,s)\right\},
\een
where $G_R^\prime$ and ${\ovl G}_R^\prime$ are functions of $p+k$, while
$G_R$ and ${\ovl G}_R$ are functions of $k$.
${\ovl G}_R$ is obtained from eq.(\ref{eq:GR}) with $m=-1$.
The fermion contribution is evaluated as 
\ben
G_R^\prime(s,t)G_R(t,s)
&=&\sum_{s,t}\left[A^\prime e^{-\alpha^\prime\vert s-t\vert}
+A^\prime_L e^{\alpha^\prime(s+t-2 N_s)}
+A^\prime_R e^{\alpha^\prime(-s-t+2)}\right]\nn\\
&&\times\left[A e^{-\alpha\vert s-t\vert}
+A_L e^{\alpha(s+t-2 N_s)}
+A_R e^{\alpha(-s-t+2)}\right]  \\
&=&A^\prime A\left\{\frac{-2e^{-(\alpha^\prime+\alpha)}}
{\left[1-e^{-(\alpha^\prime+\alpha)}\right]^2}
+N_s\frac{1+e^{-(\alpha^\prime+\alpha)}}
{1-e^{-(\alpha^\prime+\alpha)}}\right\}\nn\\
&&+A^\prime (A_L+A_R)\frac{1+e^{(\alpha^\prime+\alpha)}}
{1-e^{(\alpha^\prime+\alpha)}}\frac{e^{2\alpha}}
{1-e^{2\alpha}}\nn\\
&&+(A_L^\prime+A_R^\prime)A\frac{1+e^{(\alpha^\prime+\alpha)}}
{1-e^{(\alpha^\prime+\alpha)}}\frac{e^{2\alpha^\prime}}
{1-e^{2\alpha^\prime}}\nn\\
&&+(A_L^\prime A_L+A_R^\prime A_R)\frac{e^{2(\alpha^\prime+\alpha)}}
{\left[1-e^{(\alpha^\prime+\alpha)}\right]^2}
\een
with
\ben
A_L&=&-\frac{1-We^{-\alpha}}{1-We^{\alpha}}A,\\
A_R&=&-e^{-2\alpha}A,
\een 
where the superscript $\prime$ denotes the function of $p+k$.
On the other hand, the Pauli-Villars part is given by
\ben
{\ovl G}_R^\prime(s,t){\ovl G}_R(t,s)
&=&\sum_{s,t}A^\prime\left[e^{-\alpha^\prime\vert s-t\vert}
-e^{\alpha^\prime(s-t-N_s)}
-e^{\alpha^\prime(-s+t-N_s)}\right]\nn\\
&&\times A\left[e^{-\alpha\vert s-t\vert}
-e^{\alpha(s-t-N_s)}
-e^{\alpha(-s+t-N_s)}\right]\\
&=&A^\prime A\left\{\frac{-2e^{-(\alpha^\prime+\alpha)}}
{\left[1-e^{-(\alpha^\prime+\alpha)}\right]^2}
+N_s\frac{1+e^{-(\alpha^\prime+\alpha)}}
{1-e^{-(\alpha^\prime+\alpha)}}
+\frac{2e^{-(\alpha^\prime+\alpha)}}
{\left[1-e^{-(\alpha^\prime+\alpha)}\right]^2}
\right\},
\een
where we find the exactly same $O(N_s)$ term as 
in the domain-wall fermion part.
This is sustained even after taking the derivative in terms of $p$
according to eq.(\ref{eq:Piexpansion}). 
Through this kind of analytical calculation we have checked that 
all the $O(N_s)$ terms of the fermion contribution 
in the last term of eq.~(\ref{eq:Piexpansion})
are canceled out by the Pauli-Villars one, 
which is proportional to $N_s$ after collecting
all the terms. 

After all, eq.~(\ref{eq:Piexpansion}) is summarized as
\ben
\Pi_{\mu\nu}^{AB}(p) 
&=& - N_f\frac{g_0^2}{(4\pi)^2}\delta_{AB}\left[
(\delta_{\mu\nu}p^2-p_\mu p_\nu)
\left\{\frac{2}{3}\log\left(\frac{\pi^2}{p^2a^2}\right)
+\frac{5}{9}\right\}+\delta_{\mu\nu}p^2/3
\right.\nn\\ 
&&\left. +A\delta_{\mu\nu}p^2+B p_\mu p_\nu\right].
\label{eq:Pi_summary}
\een
The momentum integration is carried out numerically 
employing the Monte Carlo routine BASES\cite{bases}.
We choose $\Lambda=\pi$ for the cut-off. 
The results are checked by a mode sum for a periodic box of
a size $L^4$ with $L=128$ after transforming the momentum 
variable through $k^\prime_\mu=k_\mu-{\rm sin}k_\mu$.
The estimated values for $A+1/3$ and $B$ are presented in Table~\ref{tab:vp}
as a function of $M$. We observe that the expected relation 
$A+1/3=-B$ is well satisfied within error bars.
Finally we obtain 
\be
\Pi_{\mu\nu}^{AB}(p) =-N_f\frac{g_0^2}{(4\pi)^2}\delta_{AB}
(\delta_{\mu\nu}p^2-p_\mu p_\nu)
\left[-\frac{2}{3} \log(p^2a^2)+\Delta_{\rm latt}\right],
\label{eq:vp_latt}
\ee
where 
\be
\Delta_{\rm latt}=A+\frac{8}{9}+\frac{2}{3}\log(\pi^2).
\ee
Comparing the above expression to the continuum counterpart 
in eq.(\ref{eq:vp_cont}),
we obtain the finite renormalization constant $d_f$ in eq.(\ref{eq:g2}),
\be
d_f=\frac{1}{16\pi^2}\left(\Delta_{\rm latt}-\Delta_{\rm cont}\right),
\ee 
where the numerical values are given in Table~\ref{tab:vp} 
as a function of $M$.
We find that the sign is opposite to and the magnitude is compatible 
with the Wilson, the clover and the Kogut-Susskind cases : 
$d_f=0.0066949$\cite{df_w_w,df_w_kns}, $d_f=0.0314917$\cite{df_sw}, and
$d_f=0.0026248$\cite{df_ks}, respectively.
   
\section{The ratio of continuum and lattice $\Lambda$ parameters}

We first show how to obtain the mean-field improved $\msbar$ 
coupling $g_\msbar^2(\mu )$ at the scale $\mu$ from the lattice
bare coupling $g_0^2$. A detailed description for the quenched case
is already given in Sec.~VI of Ref.~\cite{pt_rg}.
For the full QCD case the mean field improvement modifies 
the relation of $g_\msbar^2(\mu )$ and $g_0^2$ in eq.(\ref{eq:g2}):  
\ben
\dfrac{1}{g_{\overline{\rm MS}}^2(\mu )}
&=& \dfrac{P}{g^2_0} + d_g + c_p +\dfrac{22}{16\pi^2} \log (\mu a)
+N_f\left(d_f -\dfrac{4}{48\pi^2} \log (\mu a)\right) .
\label{eq:g2_plaq}
\een
One may use an alternative formula\cite{cppacs}
\ben
\dfrac{1}{g_{\overline{\rm MS}}^2(\mu )}
&=& \dfrac{c_0 P + 8 c_1 R1+ 16c_2 R2 +8 c_3 R3}{g^2_0} \nn \\
& &+ d_g + (c_0\cdot c_p + 8 c_1\cdot c_{R1}+16 c_2\cdot c_{R2}+
8 c_3\cdot c_{R3})
+\dfrac{22}{16\pi^2} \log (\mu a) \nn\\
&&+N_f\left(d_f -\dfrac{4}{48\pi^2} \log (\mu a)\right),
\label{eq:g2_rg}
\een
where
\ben
P  &=& \frac{1}{3}{\rm Tr} U_{plaquette}    =1 - c_{p} g_0^2 +O(g_0^4),\\ 
R1 &=& \frac{1}{3}{\rm Tr} U_{rectangle}    =1 - c_{R1} g_0^2 +O(g_0^4),\\
R2 &=& \frac{1}{3}{\rm Tr} U_{chair}        =1 - c_{R2} g_0^2 +O(g_0^4),\\
R3 &=& \frac{1}{3}{\rm Tr} U_{parallelogram}=1 - c_{R3} g_0^2 +O(g_0^4),
\een
and the measured values may be employed for $P$, $R1$, $R2$ and $R3$. 
Table~\ref{tab:MF} summarizes 
the values of $c_{p}$, $c_{R1}$, $c_{R2}$ and $c_{R3}$ for various 
gauge actions, which are taken from Table~XVI of Ref.~\cite{pt_rg}.
One may also use the mean-field estimate for $M$, which is given by
$\tilde M = M-4(1-P^{1/4})$\cite{pt_rg}. No other modifications are necessary
in this case: Just use $d_f$ at $\tilde M$.

As discussed in Sec.~\ref{sec:intro}, 
we can determine the ratio of $\Lambda_L$ and $\Lambda_\msbar$ from the
finite renormalization of the coupling constant.
 From eq.(\ref{eq:ratio}), the ratio is written as
\ben
\frac{\Lambda_\msbar}{\Lambda_L}
&=&{\rm exp}\left(-\frac{d_g}{2\beta_0}\right)
\times{\rm exp}\left(-\frac{N_f d_f}{2\beta_0}\right),
\label{eq:ratio_gf}
\een
where the first factor is given in Table~\ref{tab:MF} for the
various gauge actions, and the second one is given  
in Table~\ref{tab:vp} as a function of $M$, in the case that $N_f=0,1,2,3,4$.

\section{Conclusion}

In this paper we evaluate the fermion contribution to the 
finite renormalization of the coupling constant in the domain-wall QCD, 
which determines the ratio of the lattice $\Lambda$ parameter 
to the continuum one. 
The numerical values are given as a function of $M$. 
We show that the massive fermion modes emerging from 
the fermion loop contribution to the gluon self energy
can be subtracted by the Pauli-Villars regulator.
We explain how to implement the mean field improvement 
for the renormalized coupling 
$g^2_\msbar(\mu)$ and the $\Lambda$ parameter ratio.

\section*{Acknowledgments}
This work is supported in part by the Grants-in-Aid for
Scientific Research from the Ministry of Education, 
Culture, Sports, Science and Technology.
(Nos. 13135204, 14046202, 15204015, 15540251, 15740165).

\input{ref.tex}

\begin{sidetable}
\caption{Values of the perturbative quantities $d_g$, $c_P$, and
$c_{Ri}$ for the various gauge actions.
Gluon contribution to the $\Lambda$ parameter ratio 
in eq.(\protect{\ref{eq:ratio_gf}}) 
is given for the case of $N_f=0,1,2,3,4$.}
\label{tab:MF}
\begin{center}
\begin{tabular}{lll|lllll|lllll}
&&&&&&&& \multicolumn{5}{|c}{${\rm exp}(-d_g/(2\beta_0))$}  \cr
action & $c_1$ & $c_3$ &  $d_g$ & $c_p$ & $c_{R1}$& $c_{R2}$ & $c_{R3}$ &
$N_f=0$ & $N_f=1$ & $N_f=2$ & $N_f=3$ & $N_f=4$  \cr
\hline
Plaquette& 0       & 0    & $-$0.4682 & 1/3   & 0.5747 & 0.5228 & 0.5687 &
28.81(2) & 35.78(3) & 45.80(4) & 60.80(5) & 84.45(8) \cr
Iwasaki  & $-$0.331  & 0    & 0.1053 & 0.1401 & 0.2689 & 0.2223 & 0.2465 &
0.4696(3) & 0.4473(3) & 0.4231(3) & 0.3970(3) & 0.3687(3) \cr
DBW2     & $-$1.40686& 0    & 0.5317 & 0.0511 & 0.1040 & 0.0818 & 0.0921 &
0.02200(2) & 0.01720(1) & 0.01300(1) &  0.009423(8) &  0.006488(6) \cr
Symanzik & $-$1/12   & 0    & $-$0.2361 & 0.2440 & 0.4417 & 0.3846 & 0.4215 &
5.445(4) & 6.074(5) & 6.879(6) & 7.935(7) & 9.365(9) \cr
Iwasaki' & $-$0.27   & $-$0.04&        & 0.1471  & 0.2797 & 0.2342 & 0.2611 &
 &  &  &  &  \cr
Wilson   & $-$0.252  & $-$0.17& 0.1196 & 0.1286 & 0.2439 & 0.2070 & 0.2352 &
0.4238(3) & 0.4010(3) & 0.3765(3) & 0.3502(3) & 0.3220(3) \cr
\end{tabular}
\end{center}
\end{sidetable}

\newpage

\begin{table}[h]
\caption{Numerical values for $A+1/3$, $B$ and $d_f$ as a function of $M$
Fermion contribution to the $\Lambda$ parameter ratio 
in eq.(\protect{\ref{eq:ratio_gf}}) 
is given for the case of $N_f=1,2,3,4$.}
\label{tab:vp}
\vspace{-3mm}
\renewcommand{\arraystretch}{0.5}
\begin{center}
\begin{tabular}{l|lll|llll}
&&&&\multicolumn{4}{c}{${\rm exp}(-N_f d_f/(2\beta_0))$}\\
$M$ & $A+1/3$ & $B$ & $d_f$ &
$N_f=1$ & $N_f=2$ & $N_f=3$ & $N_f=4$  \\
\hline
0.05 & $-$5.4253(27)  & 5.4259(27)  & $-$0.028209(17)  & 1.24054(16)  & 1.58537(44)  & 2.10102(95)  & 2.9127(19) \\
0.10 & $-$4.6075(22)  & 4.6092(23)  & $-$0.023030(14)  & 1.19240(13)  & 1.45676(33)  & 1.83331(68)  & 2.3937(13) \\ 
0.15 & $-$4.1408(20)  & 4.1414(19)  & $-$0.020075(13)  & 1.16578(11)  & 1.38810(29)  & 1.69612(57)  & 2.1400(10) \\    
0.20 & $-$3.8102(18)  & 3.8106(18)  & $-$0.017981(11)  & 1.14728(10)  & 1.34143(25)  & 1.60520(48)  & 1.97679(85) \\
0.25 & $-$3.5497(18)  & 3.5494(18)  & $-$0.016332(11)  & 1.132911(97) & 1.30577(24)  & 1.53700(45)  & 1.85699(79) \\ 
0.30 & $-$3.3341(15)  & 3.3346(16)  & $-$0.0149658(97) & 1.121148(83) & 1.27696(20)  & 1.48273(38)  & 1.76331(65) \\ 
0.35 & $-$3.1475(15)  & 3.1483(14)  & $-$0.0137846(96) & 1.111075(81) & 1.25255(20)  & 1.43734(36)  & 1.68612(61) \\ 
0.40 & $-$2.9819(14)  & 2.9825(13)  & $-$0.0127360(87) & 1.102208(73) & 1.23128(17)  & 1.39822(32)  & 1.62042(53) \\ 
0.45 & $-$2.8310(13)  & 2.8316(12)  & $-$0.0117805(83) & 1.094190(69) & 1.21221(16)  & 1.36349(30)  & 1.56279(49)  \\ 
0.50 & $-$2.6925(12)  & 2.6938(12)  & $-$0.0109029(79) & 1.086878(65) & 1.19496(15)  & 1.33236(28)  & 1.51167(45) \\ 
0.55 & $-$2.5635(12)  & 2.5650(12)  & $-$0.0100864(76) & 1.080117(62) & 1.17912(15)  & 1.30403(26)  & 1.46560(42) \\  
0.60 & $-$2.4418(12)  & 2.4433(11)  & $-$0.0093153(73) & 1.073772(60) & 1.16436(14)  & 1.27784(25)  & 1.42339(40) \\  
0.65 & $-$2.3272(11)  & 2.3282(11)  & $-$0.0085901(67) & 1.067839(55) & 1.15064(13)  & 1.25368(22)  & 1.38480(35) \\ 
0.70 & $-$2.2176(10)  & 2.2182(11)  & $-$0.0078955(64) & 1.062186(52) & 1.13767(12)  & 1.23097(21)  & 1.34883(33) \\  
0.75 & $-$2.11237(99) & 2.11208(98) & $-$0.0072294(63) & 1.056794(51) & 1.12535(12)  & 1.20958(20)  & 1.31520(31) \\
0.80 & $-$2.01129(98) & 2.01132(97) & $-$0.0065893(62) & 1.051638(50) & 1.11365(11)  & 1.18937(19)  & 1.28368(30) \\  
0.85 & $-$1.91395(88) & 1.91450(89) & $-$0.0059729(55) & 1.046696(44) & 1.10249(10)  & 1.17023(17)  & 1.25404(26) \\ 
0.90 & $-$1.82014(87) & 1.82057(91) & $-$0.0053788(55) & 1.041956(44) & 1.091844(98) & 1.15207(17)  & 1.22612(26) \\ 
0.95 & $-$1.73060(81) & 1.73063(86) & $-$0.0048118(51) & 1.037451(40) & 1.081777(90) & 1.13501(15)  & 1.20005(23) \\ 
1.00 & $-$1.64359(81) & 1.64365(82) & $-$0.0042608(51) & 1.033092(41) & 1.072083(90) & 1.11867(15)  & 1.17525(23) \\  
1.05 & $-$1.56012(77) & 1.55976(76) & $-$0.0037322(49) & 1.028928(39) & 1.062866(85) & 1.10321(14)  & 1.15194(21) \\ 
1.10 & $-$1.48027(71) & 1.47976(71) & $-$0.0032266(45) & 1.024961(35) & 1.054123(78) & 1.08863(13)  & 1.13008(19) \\ 
1.15 & $-$1.40379(68) & 1.40313(69) & $-$0.0027422(43) & 1.021174(33) & 1.045815(73) & 1.07484(12)  & 1.10952(18) \\ 
1.20 & $-$1.33232(65) & 1.33289(66) & $-$0.0022896(41) & 1.017649(32) & 1.038112(70) & 1.06211(12)  & 1.09065(17) \\ 
1.25 & $-$1.26633(62) & 1.26646(62) & $-$0.0018718(39) & 1.014405(30) & 1.031050(66) & 1.05050(11)  & 1.07352(16) \\ 
1.30 & $-$1.20519(59) & 1.20600(60) & $-$0.0014846(37) & 1.011408(29) & 1.024549(63) & 1.03985(10)  & 1.05788(15) \\ 
1.35 & $-$1.15142(57) & 1.15256(57) & $-$0.0011441(36) & 1.008780(28) & 1.018865(60) & 1.030569(98) & 1.04431(14) \\ 
1.40 & $-$1.10734(55) & 1.10903(55) & $-$0.0008650(35) & 1.006631(27) & 1.014230(58) & 1.023026(94) & 1.03332(14) \\ 
1.45 & $-$1.07434(53) & 1.07514(54) & $-$0.0006560(34) & 1.005025(26) & 1.010774(56) & 1.017416(90) & 1.02517(13) \\ 
1.50 & $-$1.05539(52) & 1.05533(52) & $-$0.0005360(33) & 1.004104(25) & 1.008794(55) & 1.014206(89) & 1.02052(13) \\ 
1.55 & $-$1.05344(53) & 1.05437(53) & $-$0.0005236(33) & 1.004009(26) & 1.008590(55) & 1.013876(89) & 1.02004(13) \\ 
1.60 & $-$1.07444(54) & 1.07497(54) & $-$0.0006566(34) & 1.005030(26) & 1.010784(56) & 1.017431(91) & 1.02520(13) \\ 
1.65 & $-$1.12617(56) & 1.12665(56) & $-$0.0009842(36) & 1.007548(27) & 1.016207(59) & 1.026241(96) & 1.03800(14) \\ 
1.70 & $-$1.21873(61) & 1.21856(61) & $-$0.0015704(39) & 1.012071(30) & 1.025985(65) & 1.04220(11)  & 1.06132(16) \\ 
1.75 & $-$1.36869(68) & 1.36952(68) & $-$0.0025200(43) & 1.019442(34) & 1.042025(74) & 1.06857(12)  & 1.10022(18) \\ 
1.80 & $-$1.60462(80) & 1.60601(80) & $-$0.0040140(51) & 1.031146(40) & 1.067771(89) & 1.11143(15)  & 1.16431(22) \\ 
1.85 & $-$1.97527(99) & 1.97602(99) & $-$0.0063612(63) & 1.049807(50) & 1.10951(11)  & 1.18225(19)  & 1.27263(30) \\ 
1.90 & $-$2.5818(13)  & 2.5816(13)  & $-$0.0102019(82) & 1.081071(68) & 1.18135(16)  & 1.30800(28)  & 1.47203(46) \\ 
1.95 & $-$3.7162(19)  & 3.7153(19)  & $-$0.017386(12)  & 1.14207(10)  & 1.32845(26)  & 1.58024(49)  & 1.93268(86) \\
\end{tabular}
\end{center}
\end{table}

\clearpage 

\end{document}

%% file: ref.tex
\newcommand{\J}[4]{{ #1} {\bf #2} (#3) #4}
\newcommand{\MPL}{Mod.~Phys.~Lett.}
\newcommand{\IJMP}{Int.~J.~Mod.~Phys.}
\newcommand{\NP}{Nucl.~Phys.}
\newcommand{\PL}{Phys.~Lett.}
\newcommand{\PR}{Phys.~Rev.}
\newcommand{\PRL}{Phys.~Rev.~Lett.}
\newcommand{\AP}{Ann.~Phys.}
\newcommand{\CMP}{Commun.~Math.~Phys.}
\newcommand{\PTP}{Prog. Theor. Phys.}
\newcommand{\Suppl}{Prog. Theor. Phys. Suppl.}
\bibliography{basename of .bib file}

%% file: ms.bbl
\begin{thebibliography}{99}
\bibitem{shamir}  Y.~Shamir, Nucl. Phys. {\bf B406} (1993) 90.
\bibitem{FS} V.~Furman and Y.~Shamir, Nucl. Phys. {\bf B439} (1995) 54.
\bibitem{dw_full} RBC Collaboration, T.~Izubuchi {\it et al}.,
hep-lat/0210011 
\bibitem{dg_plaq} A.~Hasenfratz and P.~Hasenfratz, 
Phys.~Lett.~{\bf 93B} (1980) 165; Nucl.~Phys.~{\bf B193} (1981) 210. 
\bibitem{Weisz83}
 P. Weisz, \J{\NP}{B212}{1983}{1};
 P. Weisz and R. Wohlert, \J{\NP}{B236}{1984}{397};
 erratum, \J{\it ibid.}{B347}{1984}{544}.
\bibitem{dg_w} Y.~Iwasaki and S.~Sakai,
\J{\NP}{B248}{1984}{441}.
\bibitem{dg_iwa} Y.~Iwasaki and T.~Yoshi\'{e},
\J{\PL}{143B}{1984}{449}.
\bibitem{dg_dbw2} S.~Sakai,T.~Saito and A.~Nakamura,
\J{\NP}{B584}{2000}{528}.
\bibitem{pt_at}  S.~Aoki and Y.~Taniguchi,
Phys. Rev. {\bf D59} (1999) 054510.
\bibitem{pt_awi}  S.~Aoki and Y.~Taniguchi,
Phys. Rev. {\bf D59} (1999) 094506.
\bibitem{pt_2} S.~Aoki, T.~Izubuchi, Y.~Kuramashi and Y.~Taniguchi,
Phys. Rev. {\bf D59} (1999) 094505.
\bibitem{pt_34} S.~Aoki, T.~Izubuchi, Y.~Kuramashi and Y.~Taniguchi,
Phys. Rev. {\bf D60} (1999) 114504. 
\bibitem{pt_pngn} S.~Aoki and Y.~Kuramashi,
Phys. Rev. {\bf D63} (2001) 054504.
\bibitem{pt_rg} S.~Aoki, T.~Izubuchi, Y.~Kuramashi and Y.~Taniguchi,
Phys.~Rev.~{\bf D67} (2003) 094502.
\bibitem{vranas} P.~Vranas, Phys. Rev. {\bf D57} (1998) 1415.
\bibitem{LW} M.~L\"uscher and P.~Weisz, \J{\CMP}{97}{1985}{59};
erratum, {\it ibid.} {\bf 98} (1985) 433.
\bibitem{Iwasaki83}
Y.~Iwasaki,
 preprint, UTHEP-118 (Dec. 1983), unpublished.
\bibitem{Wilson}
K. G. Wilson, in {\it Recent Development of Gauge Theories},
 eds. G. 'tHooft {\it et al}. (Plenum, New York, 1980).
\bibitem{qcdtaro}T.~Takaishi, \J{\PR}{D54}{1996}{1050};
P.~de Forcrand {\it et al.}, \J{\NP}{B577}{2000}{263}.
\bibitem{cppacs_rg} CP-PACS Collaboration, A.~Ali Khan {\it et al.}, 
Phys.~Rev.~{\bf D63} (2001) 114504; 
Phys.~Rev.~{\bf D64} (2001) 114506; 
CP-PACS Collaboration, J.~I.~Noaki {\it et al.}, hep-lat/0108013. 
\bibitem{rbc_rg} RBC Collaboration, Y.~Aoki {\it et al.}, 
Nucl.~Phys.~Proc.~Suppl. {\bf 106} (2002) 245; 
RBC Collaboration, K.~Orginos {\it et al.}, 
Nucl.~Phys.~Proc.~Suppl. {\bf 106} (2002) 721; 
Y.~Aoki {\it et al.}, hep-lat/0211023. 
\bibitem{bases} S.~Kawabata, Comput.~Phys.~Commun. {\bf 41} (1986) 127;
{\bf 88} (1995) 309.
\bibitem{df_w_w} P.~Weisz,
Phys.~Lett. {\bf 100B} (1981) 331.
\bibitem{df_w_kns} H.~Kawai, R.~Nakayama and K.~Seo,
Nucl. Phys. {\bf B189} (1981) 40.
\bibitem{df_sw} S.~Sint and R.~Sommer, 
Nucl. Phys. {\bf B465} (1996) 71. 
\bibitem{df_ks} H.~S.~Sharatchandra, H.~J.~Thun and P.~Weisz,
Nucl. Phys. {\bf B192} (1981) 205. 
\bibitem{cppacs} A.~Ali Khan {\it et al.},
Phys. Rev. {\bf D65} (2002) 054505.




\end{thebibliography}
